\documentclass[prd,aps,showpacs,nofootinbib,amsmath]{revtex4}
\usepackage{epsf,epsfig,graphicx}

\begin{document}

\title{Three-parton contribution to the $B\to\pi$ form factors in $k_T$ factorization}

\author{Yu-Chun Chen}
\email{ycchen@phys.sinica.edu.tw}
\author{Hsiang-nan Li}
\email{hnli@phys.sinica.edu.tw}

\affiliation{Institute of Physics, Academia Sinica, Taipei, Taiwan
115, Republic of China} \affiliation{Department of Physics,
Tsing-Hua University, Hsinchu, Taiwan 300, Republic of China}
\affiliation{Department of Physics, National Cheng-Kung University,\\
Tainan, Taiwan 701, Republic of China}

\begin{abstract}

We calculate the three-parton twist-3 contribution to the $B\to\pi$
transition form factors in the $k_T$ factorization theorem. Since
different mesons are involved in the initial and final states,
two(three)-parton-to-three(two)-parton amplitudes do not vanish.
It is found that the dominant contribution arises from the
diagrams with the additional valence gluon attaching to the
leading-order hard gluon. Employing the three-parton
meson distribution amplitudes from QCD sum rules, we show that
this subleading piece amounts only up to few percents of the form
factors at large recoil of the pion. The
framework for analyzing three-parton contributions to $B$ meson decays
in the $k_T$ factorization is established.

\end{abstract}


\maketitle

\section{INTRODUCTION}

The $k_T$ factorization theorem is a theoretical framework
appropriate for QCD processes dominated by dynamics at small parton
momenta \cite{CCH,CE,LRS,BS,LS,HS}. With continuous efforts,
progress has been made in the application of the $k_T$ factorization
to exclusive processes at subleading level: two-parton twist-3
contributions to the pion form factor and to the $B$ meson
transition form factors have been analyzed in \cite{CKO09} and in
\cite{TLS,HQW06,Akeroyd:2007fy,Hsu:2007qc}, respectively. These
contributions are formally power-suppressed, but numerically crucial
for accommodating experimental data or lattice QCD results.
Corrections to the pion transition form factor and to the pion form
factor at next-to-leading order (NLO) in the coupling constant
were calculated in \cite{NL07} and in \cite{LWS10},
respectively. These NLO pieces can be minimized by choosing a
factorization scale close to virtuality of internal particles, and
found to be few percents in the former \cite{NL09} and about 30\% in
the latter \cite{LWS10}. The three-parton twist-3 contribution to the
pion form factor was first formulated and evaluated in the $k_T$
factorization in \cite{CL11}, and the smallness of this subleading
piece (about few percents) was confirmed.

In this paper we shall compute the three-parton twist-3 contribution to
the $B\to\pi$ transition form factors, which is down by a power of
$1/m_B$, $m_B$ being the $B$ meson mass. We stress that there are
many sources of $1/m_B$ corrections. The two-parton twist-3 one
mentioned above is suppressed by $m_0/m_B$ with the chiral scale
$m_0\approx 1.4$ GeV. It is the reason why this piece is numerically
important, namely, of the same order as the leading-twist one.
Another sizable source arises from the difference between
the two leading-twist $B$ meson distribution amplitudes
\cite{HQW06,LY03,WHF07}, which can contribute about 30\% of the form 
factors. Other power-suppressed pieces are of
order $\Lambda_{\rm QCD}/m_B$, $\Lambda_{\rm QCD}$ being the QCD
scale, and should be negligible. The $B$ meson distribution
amplitudes from higher-twist spin projectors and associated with the
three-parton Fock states belong to this category. We shall demonstrate 
that the three-parton contribution is only few percents of the
$B\to\pi$ transition form factors, consistent with the observation
made in the light-cone QCD sum rules \cite{K93}.

\section{GAUGE INVARIANCE}

\begin{figure}[t]
\begin{center}
\includegraphics[height=2.8cm]{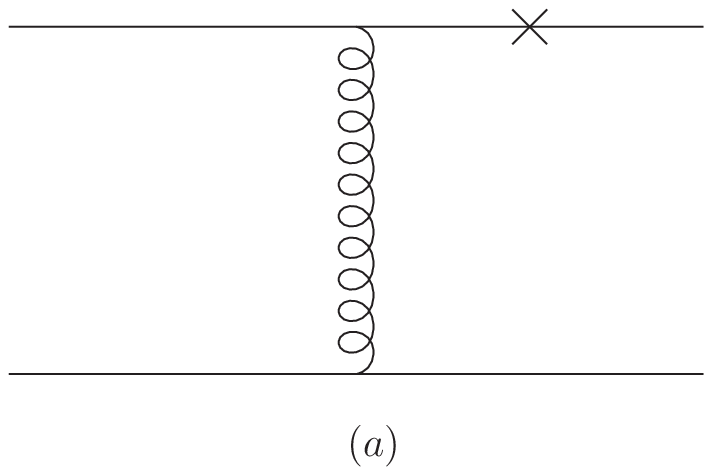}\hspace{1.0cm}
\includegraphics[height=2.8cm]{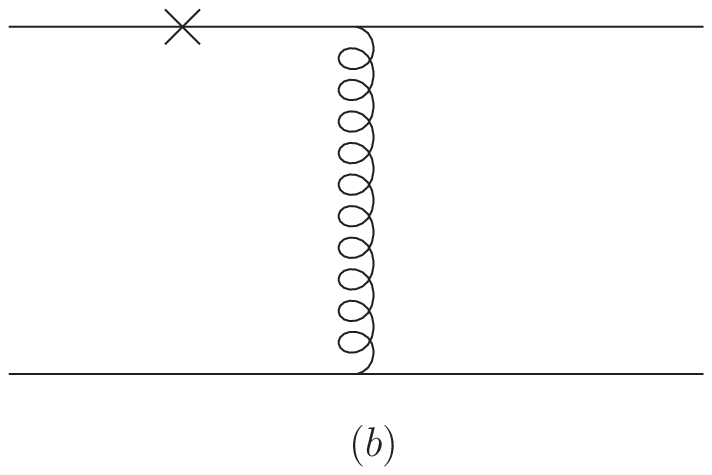}\vspace{1.0cm}
\caption{Leading-order diagrams for the $B\to\pi$ transition form
factors, where the symbol $\times$ represents the weak decay
vertex.}\label{fig1}
\end{center}
\end{figure}

\begin{figure}[t]
\begin{center}
\includegraphics[height=2.8cm]{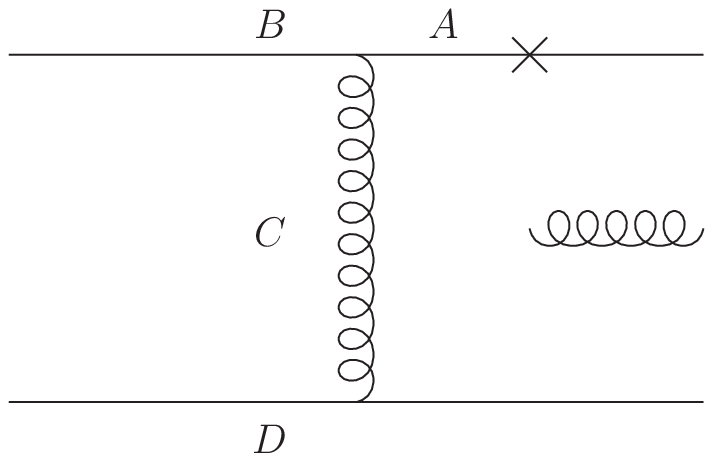}\hspace{1.0cm}
\includegraphics[height=2.8cm]{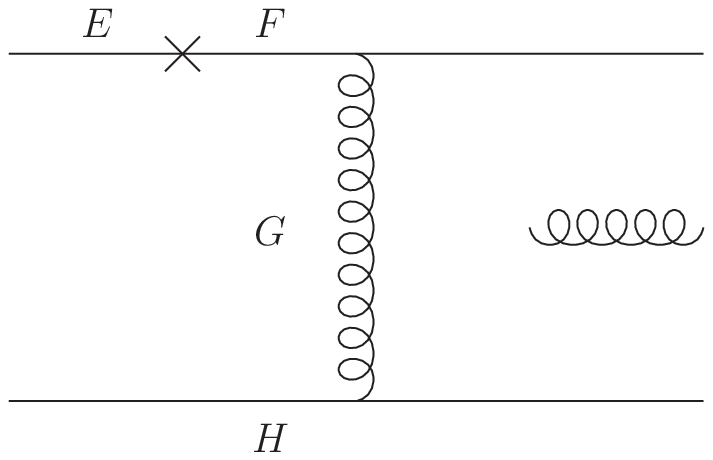}\vspace{1.0cm}
\caption{Attachments of an additional valence gluon in the
pion to lines in Fig.~\ref{fig1}.}\label{fig2}
\end{center}
\end{figure}

Compared to the collinear factorization \cite{LB80}, the
construction of the $k_T$ factorization is subtler. For example,
the gauge invariance of the $k_T$ factorization, in which
parton transverse momenta are retained, becomes an issue \cite{LM09}.
The gauge invariance of the $k_T$ factorization for the $B\to\pi$
transition form factors at the three-parton twist-3 level can be proved
in a way similar to the case of the pion form factor \cite{CL11}.
We display in Fig.~\ref{fig1} the leading-order (LO) diagrams,
and in Fig.~\ref{fig2} the attachments of an additional valence gluon from
the pion to all the lines in the LO diagrams, except the valence
quark lines in the pion. There are two sources of gauge dependence
\cite{Q90}, which arise from the patron transverse
momentum in Fig.~\ref{fig1} and from the three-parton Fock state in
Fig.~\ref{fig2}. We shall show in this section that the
gauge-dependent amplitudes cancel in each of the two sources.
The proof for the amplitudes with three partons from the $B$ meson side is
the same.

The $B$ meson momentum $P_1$ and the pion momentum $P_2$ are
parameterized as
\begin{eqnarray}
P_1=(P_1^+,P_1^-,{\bf 0}_T)=\frac{m_B}{\sqrt{2}}(1,1,{\bf
0}_T),\;\;\;\; P_2=(0,P_2^-,{\bf
0}_T)=\frac{m_B}{\sqrt{2}}(0,\eta,{\bf 0}_T), \label{mpp}
\end{eqnarray}
where the energy fraction $\eta=1-q^2/m_B^2$ carried by the pion
ranges between 0 and 1. The momenta of the antiquarks in the
$B$ meson and in the pion, represented by the lower fermion line,
are parameterized as
\begin{eqnarray}
k_1=(x_1P_1^+,0,{\bf k}_{1T}),\;\;\;\; k_2=(0,x_2P_2^-,{\bf
k}_{2T}),
\end{eqnarray}
respectively, $x_1$ and $x_2$ being the momentum fractions. It is
understood that the components $k_1^-$ and $k_2^+$ have been dropped
in hard kernels, and integrated out of the $B$ meson and pion wave 
functions, respectively. The gluon propagator of momentum $l$ is written as
\begin{eqnarray}
\frac{-i}{l^2}\left(g^{\sigma\nu} - \lambda \frac{l^\sigma
l^\nu}{l^2}\right),\label{tensor}
\end{eqnarray}
in the covariant gauge, where the parameter $\lambda$ is used to
identify sources of gauge dependence.

We sandwich Fig.~\ref{fig1}(a) with the spin projectors
\begin{eqnarray}
\frac{1}{4N_c}(\not P_1 + m_B)\gamma_5,\;\;\;\;
\frac{1}{4N_c}\gamma_5\gamma_{\beta},\label{str}
\end{eqnarray}
from the initial and final states, respectively, where $N_c=3$ is
the number of colors, $\gamma_5\gamma_{\beta}$ is a higher-twist
projector \cite{CL11} selected for the proof below, and the subscript
$\beta$ takes the transverse components.
The resultant hard kernel contains the gauge-dependent piece
\begin{eqnarray}
H^{a\lambda}=\frac{1}{16}g^2 \frac{C_F}{N_c}\lambda\frac{tr[
\gamma^\sigma \gamma_5\gamma^{\beta}\gamma_\mu(1-\gamma_5) (\not
P_1-\not k_2+m_b)\gamma^\nu(\not P_1 + m_B)
\gamma_5]}{[(P_1-k_2)^2-m_b^2](k_1-k_2)^2}\frac{(k_1-k_2)_\sigma
(k_1-k_2)_\nu}{(k_1-k_2)^2},\label{hlam}
\end{eqnarray}
with the $b$ quark mass $m_b$. In the small $x$ region where the
$k_T$ factorization applies, we keep the transverse momentum
dependence in the denominator \cite{NL2}. The transverse momentum
dependence in the numerator belongs to the twist-3 contribution.
Inserting the identity $\not\! k_1 - \not\! k_2=(\not\! P_1-\not\!
k_2-m_b)-(\not\! P_1- \not\! k_1 - m_b)$ for the gluon vertex on the $b$
quark line, we find that the second term vanishes at leading-twist
accuracy on the $B$ meson side, as it is multiplied by $\not\! P_1 +
m_B$. The derivative of the numerator with respect to $k_{2T}^\beta$ gives
\begin{eqnarray}
H^{a\lambda}_{T}&\equiv &\frac{\partial }{\partial
k_{2T}^\beta}H^{a\lambda}
= -\frac{g^2}{16}\frac{C_F}{N_c}\lambda\frac{tr[
\gamma_\beta\gamma_5\gamma^\beta\gamma_\mu(1-\gamma_5)( \not
P_1+m_B)\gamma_5]}{(k_1-k_2)^4}.\label{at2}
\end{eqnarray}

The LO hard kernel from Fig.~\ref{fig1}(b) contains the
gauge-dependent amplitude
\begin{eqnarray}
H^{b\lambda}=\frac{1}{16}g^2 \frac{C_F}{N_c}\lambda\frac{tr[ \gamma^\sigma
\gamma_5\gamma^{\beta}\gamma^\nu(\not P_2-\not k_1)\gamma_\mu
(1-\gamma_5)(\not P_1 +
m_B)\gamma_5]}{(P_2-k_1)^2(k_1-k_2)^2}\frac{(k_1-k_2)_\sigma
(k_1-k_2)_\nu}{(k_1-k_2)^2}.\label{hlbm}
\end{eqnarray}
The similar differentiation with respect to $k_{2T}^\beta$ leads to
\begin{eqnarray}
H^{b\lambda}_{T}
\nonumber & = & -\frac{1}{16}g^2 \frac{C_F}{N_c}\lambda\frac{tr[ (\not k_1- \not
k_2)\gamma_5\gamma^{\beta}\gamma_{\beta}(\not P_2 - \not
k_1)\gamma_\mu(1-\gamma_5) (
\not P_1 + m_B)\gamma_5]}{(P_2-k_1)^2(k_1-k_2)^4}\\
\nonumber &  & - \frac{1}{16}g^2 \frac{C_F}{N_c}\lambda\frac{tr[ \gamma_\beta
\gamma_5\gamma^{\beta}(\not k_1 - \not k_2)(\not P_2-\not
k_1)\gamma_\mu(1-\gamma_5) (\not P_1 +
m_B)\gamma_5]}{(P_2-k_1)^2(k_1-k_2)^4}.
\end{eqnarray}
For the first term in the above expression, $\not\! k_1$ ($\not\! k_2$)
implies one more derivative of the spectator field on the $B$ meson (pion)
side, so it is neglected. The second term, after employing
$\not\! k_1 - \not\! k_2=(\not\! P_2-\not\! k_2)- (\not\! P_2-\not\! k_1)$, gives
\begin{eqnarray}
H^{b\lambda}_{T}
& =& \frac{1}{16}g^2 \frac{C_F}{N_c}\lambda\frac{tr[ \gamma_\beta
\gamma_5\gamma^{\beta}\gamma_\mu(1-\gamma_5) (\not P_1 +
m_B)\gamma_5]}{(k_1-k_2)^4},\label{bt2}
\end{eqnarray}
where the term $\not\! P_2-\not\! k_2$, implying the derivative of the energetic
quark field on the pion side, has been dropped. Apparently,
Eqs.~(\ref{at2}) and (\ref{bt2}) cancel each. That is, the gauge
dependence associated with the derivative of quark fields disappears
at $1/m_B$.

We then compute the gauge-dependent amplitudes from Fig.~\ref{fig2}, in which
any contributions from the derivatives of quark fields should be ignored.
For the attachment $A$ of the valence gluon to the virtual quark line, the spin
projector for the pion in Eq.~(\ref{str}) is replaced by $\gamma_5\gamma_{\beta}/2$
\cite{CL11}. The color factor associated with this
attachment is given by $tr[T^aT^bT^bT^c]=C_F \delta^{ac}/2$, where
the index $a$ labels the color of the valence gluon. Summing over
the index $c$, the corresponding amplitude is written as
\begin{eqnarray}
H^{\lambda}_A
&=&\frac{1}{16}g^2\frac{C_F}{N_c} \lambda\frac{tr[(\not
k_1-\not k_2) \gamma_5\gamma^{\beta}\gamma_\mu(1-\gamma_5) (\not
P_1-\not k_2-\not l_2+m_b)\gamma_\beta(\not
P_1+m_B)\gamma_5]}{[(P_1-k_2-l_2)^2-m_b^2](k_1-k_2)^4},
\end{eqnarray}
which vanishes because $\not\! k_1$ ($\not\! k_2$) in the factor $\not\!
k_1-\not\! k_2$ implies one more derivative of the spectator field on
the $B$ meson (pion) side. Using a similar argument, the attachment
$B$ does not generate the gauge dependence either.

The gauge-dependent amplitude from the attachment $C$ is given by
\begin{eqnarray}
H^{\lambda}_C&=&-\frac{1}{32}g^2 \frac{tr[
\gamma_{\delta'}\gamma_5\gamma^{\beta}\gamma_\mu(1-\gamma_5) (\not
P_1-\not k_2-\not l_2 + m_b)\gamma_{\nu'}(\not
P_1+m_B)\gamma_5]}{[(P_1-k_2-l_2)^2-m_b^2]}
\nonumber\\
& &\times [g_{\beta\nu}(2l_2-k_1+k_2)_\delta
+g_{\nu\delta}(-2k_2+2k_1-l_2)_\beta
+g_{\delta\beta}(-k_1+k_2-l_2)_\nu]\nonumber\\
& &\times\Bigg[\lambda\frac{g^{\delta\delta'}}{(k_1-k_2)^2}
\frac{(k_1-k_2-l_2)^\nu(k_1-k_2-l_2)^{\nu'}}{(k_1-k_2-l_2)^4}+\lambda
\frac{(k_1-k_2)^\delta(k_1-k_2)^{\delta'}}{(k_1-k_2)^4}
\frac{g^{\nu\nu'}}{(k_1-k_2-l_2)^2}\nonumber\\
& &-\lambda^2\frac{(k_1-k_2)^\delta(k_1-k_2)^{\delta'}}{(k_1-k_2)^4}
\frac{(k_1-k_2-l_2)^\nu(k_1-k_2-l_2)^{\nu'}}{(k_1-k_2-l_2)^4}\Bigg].
\end{eqnarray}
According to the above explanation, if the vertex on the spectator
line contains $\not\! k_1-\not\! k_2$, the associated term comes from the
derivative of the spectator field, and should be dropped. Hence, the
gauge dependence can appear only in the first term linear in
$\lambda$, which leads to
\begin{eqnarray}
H^{\lambda}_C
& = & \frac{1}{32}\lambda g^2 \frac{tr[
\gamma_{\beta} \gamma_5\gamma^{\beta} \gamma_\mu(1-\gamma_5)(\not
P_1+m_B)\gamma_5]}{(k_1-k_2-l_2)^4}.\label{ac}
\end{eqnarray}

The evaluation of the gauge-dependent pieces for the rest of
attachments is similar. With the color factor for the
attachment $D$, $tr[T^bT^aT^bT^c]=-\delta^{ac}/(4N_c)$,
the corresponding amplitude is written as
\begin{eqnarray}
H^{\lambda}_D
&=&-\frac{1}{32}g^2\frac{1}{N_c^2}\lambda\frac{tr[\gamma_\beta\gamma_5\gamma^{\beta}\gamma_\mu(1-\gamma_5)
(\not P_1+m_B)\gamma_5]}{(k_1-k_2-l_2)^4}.\label{ad}
\end{eqnarray}
The gauge-dependent amplitudes from the attachments $E$ and $F$ diminish.
The attachments $G$ and $H$ give
\begin{eqnarray}
H^{\lambda}_G
&=&-\frac{1}{32}\lambda g^2
\frac{tr[ \gamma^{\beta} \gamma_5\gamma_{\beta} \gamma_\mu(1-\gamma_5)(\not
P_1+m_B)\gamma_5]}{(k_1-k_2-l_2)^4},\label{bg}\\
H^{\lambda}_H
&=& \frac{1}{32}g^2\frac{1}{N_c^2}
\lambda\frac{tr[\gamma_\beta\gamma_5\gamma^{\beta}\gamma_\mu(1-\gamma_5)(\not
P_1+m_B)\gamma_5]}{(k_1-k_2-l_2)^4},\label{bh}
\end{eqnarray}
respectively. The cancellation between Eqs.~(\ref{ac}) and (\ref{bg}), and between
Eqs.~(\ref{ad}) and (\ref{bh}) is observed. That is, the gauge
dependence from the three-parton Fock state also disappears. This
completes the proof of the gauge invariance of the $k_T$
factorization for the $B\to\pi$ transition form factors at leading
order in $\alpha_s$ and at three-parton twist-3 level.

\section{THREE-PARTON CONTRIBUTIONS}

In this section we calculate the $B\to\pi$ transition form factors
$F_+$ and $F_0$ involved in the semileptonic decay $B(P_1)\to\pi(P_2)\ell\nu$,
\begin{eqnarray}
\langle\pi(P_2)|{\bar b}(0)\gamma_\mu u(0)|B(P_1)\rangle
=F_+(q^2)\left[(P_1+P_2)_\mu-\frac{m_B^2}{q^2}q_\mu\right]
+F_0(q^2)\frac{m_B^2}{q^2}q_\mu,
\end{eqnarray}
where $q=P_1-P_2$ is the lepton-pair momentum. Another equivalent
definition is given by
\begin{eqnarray}
\langle\pi(P_2)|{\bar b}(0)\gamma_\mu u(0)|B(P_1)\rangle
=f_1(q^2)P_{1\mu}+f_2(q^2)P_{2\mu},
\end{eqnarray}
in which the form factors $f_1$ and $f_2$ are related to $F_+$ and
$F_0$ via
\begin{eqnarray}
F_+=\frac{f_1+f_2}{2},\;\;\;\;
F_0=\frac{f_1}{2}\left(1+ \frac{q^2}{m_B^2}\right)
+\frac{f_2}{2}\left(1 - \frac{q^2}{m_B^2}\right).\label{f+0}
\end{eqnarray}

We start with
the hard kernels from the two-parton-to-three-parton diagrams in the
Feynman gauge ($\lambda=0$).  
The following matrix element \cite{EKT06} defines
the three-parton twist-3 pion wave function $T(z,z')$,
\begin{eqnarray}
\langle 0|{\bar q}(z)\sigma^+_{\hspace{0.2cm}\alpha'}\gamma_5
gG^+_{\hspace{0.2cm}\alpha}(z') q(0) |\pi(P_1)\rangle=if_\pi
m_0(P_1^+)^2g^T_{\alpha\alpha'}T(z,z'),\label{gi3}
\end{eqnarray}
with the chiral scale $m_0=m_\pi^2/(m_u+m_d)$, $m_\pi$, $m_u$, and
$m_d$ being the pion, $u$ quark and $d$ quark masses, respectively.
The three momenta $P_2-k_2-l_2, k_2$, and $l_2$ are assigned 
to the final-state quark, antiquark,
and gluon, respectively. For the calculation, 
we replace the projector for the pion in Eq.~(\ref{str})
by $\gamma_5\not\! P_2\gamma^T_{\beta} m_0/(4y_2)$ \cite{CL11},
where the valence gluon momentum fraction is defined by
$y_2=l_2^-/P_2^-$, the gamma matrix $\gamma^T$ contains only
transverse components, and the pion decay constant has been absorbed
into the wave function $T(z,z')$.

The amplitudes from the attachments $A$, $B$, $\cdots$, $H$ in
Fig.~\ref{fig2} are collected as follows:
\begin{eqnarray}
H_A^{2\to 3}&=&  \frac{g^2C_F}{2 N_cy_2}\left[
\frac{1}{(P_1-k_2-l_2)^2-m_b^2} +\frac{1}{(P_1-k_2)^2-m_b^2}\right]
\frac{m_Bm_0P_{2\mu}}{(k_1-k_2)^2},\label{a23}\\
H_B^{2\to 3}&=&0,\label{b23}\\
H_C^{2\to 3}&=&  \frac{g^2}{8}\frac{\eta(x_2-y_2)-x_1}
{\eta y_2(x_2+y_2)}\frac{m_Bm_0P_{2\mu}}{(k_1-k_2)^2(k_1-k_2-l_2)^2},\label{c23}\\
H_D^{2\to 3}& = & -\frac{g^2}{4N_c^2}\frac{1}
{x_2+y_2}\frac{m_Bm_0P_{2\mu}}{(k_1-l_2)^2(k_1-k_2-l_2)^2},\\
H_E^{2\to 3}&=&H_F^{2\to 3}=H_H^{2\to 3}=0,\label{e23}\\
H_G^{2\to 3}& = & -\frac{g^2}{8y_2}\frac{m_Bm_0 (P_{2\mu}+k_{1\mu})}
{(k_1-k_2)^2(k_1-k_2-l_2)^2}.\label{g23}
\end{eqnarray}
The denominators of Eqs.~(\ref{a23}) and (\ref{c23}) indicate that the
contribution from the former is down by a power of $k_1^+/m_B\sim
\Lambda_{\rm QCD}/m_B$. That is, the attachments to the $b$ quark line and to
the energetic parton line of the pion give power-suppressed
contributions in the dominant region with soft spectator momenta.
This observation is similar to that obtained in the study of the
three-parton twist-3 contribution to the pion form factor \cite{CL11}.
Equation~(\ref{b23}) vanishes, since the $\gamma$ matrix associated
with the valence gluon attachment takes only the transverse
components. One can then flip the $b$ quark propagator and this
$\gamma$ matrix, and apply $(\not\! P_1- \not\! k_1 - m_b)(\not\! P_1 +
m_B)\approx 0$ at leading-twist accuracy on the $B$ meson side.
The attachments $E$, $F$, and $H$ do not contribute as shown
in Eq.~(\ref{e23}), simply because of
$\gamma_\nu\gamma_5\not\! P_2\gamma_{\beta}^T\gamma^\nu=0$. The
$k_{1\mu}$ term in Eq.~(\ref{g23}) is of higher-power and
negligible.

It is found that all the above amplitudes are proportional to $m_B$, namely,
diminish as $m_B\to 0$. This must
be the case, since the two(three)-parton-to-three(two)-parton
diagrams do not contribute to the pion form factor \cite{CL11}. In the
numerical analysis below we shall not differentiate $m_B$ and $m_b$,
whose difference gives an additional power of $1/m_B$.
Ignoring Eq.~(\ref{a23}) and the second term in Eq.~(\ref{g23}),
the two-parton-to-three-parton amplitudes are summed into
\begin{eqnarray}
H^{2\to 3}&=&-\frac{g^2}{8(x_2+y_2)}\left[\frac{2\eta y_2+x_1}{\eta y_2}
\frac{1}{(k_1-k_2)^2}+\frac{2}{N_c^2}\frac{1}{(k_1-l_2)^2}\right]
\frac{m_Bm_0 P_{2\mu}}{(k_1-k_2-l_2)^2}.
\end{eqnarray}
To derive the above expression, we have followed the hierarchy among
the relevant scales $xm_B^2 \gg k_T^2$ \cite{LWS10,CL11}, under which
the $k_T$-dependent terms in the denominators of the $b$
quark and energetic quark propagators are dropped.

For the three-parton-to-two-parton amplitudes, we need to introduce
the three-parton $B$ meson distribution amplitude. Consider the following
matrix elements associated with the $\bar B$ meson \cite{GN97}
\begin{eqnarray}
& &\langle 0|\bar q{\bf \alpha}\cdot g{\bf E}\gamma_5h_v|\bar B(v)\rangle
=F(\mu)\lambda_E^2(\mu),\nonumber\\
& &\langle 0|\bar q{\bf \sigma}\cdot g{\bf H}\gamma_5h_v|\bar B(v)\rangle
=iF(\mu)\lambda_H^2(\mu),\label{bt3}
\end{eqnarray}
where $E^i = G^{0i}$ and $H^i = (-1/2)\epsilon^{ijk}G^{jk}$ are the
chromoelectric and chromomagnetic fields, respectively, $h_v$ is the
effective heavy quark field, $v=(1,{\bf 0})$ is the $\bar B$ meson
velocity, and $\mu$ is the renormalization scale. The normalization
$F(\mu)=f_B\sqrt{m_B}+O(\alpha_s,1/m_B)$, $f_B$ being the $B$ meson
decay constant, is defined via
$\langle 0|\bar q\gamma_\rho\gamma_5h_v|\bar B(v)\rangle
=iF(\mu)v_\rho$.
The analysis based on QCD sum rules in \cite{GN97} and
\cite{NT11} led to the values
\begin{eqnarray}
& &\lambda_E^2(1\;\;{\rm GeV}) = (0.11 \pm 0.06)\;\;{\rm
GeV}^2,\;\;\;\; \lambda_H^2(1\;\;{\rm GeV}) = (0.18 \pm
0.07)\;\;{\rm GeV}^2,\label{neu}\\
& &\lambda_E^2(1\;\;{\rm GeV}) = (0.03 \pm 0.02)\;\;{\rm
GeV}^2,\;\;\;\; \lambda_H^2(1\;\;{\rm GeV}) = (0.06 \pm
0.03)\;\;{\rm GeV}^2,\label{tan}
\end{eqnarray}
respectively. The two matrix elements in Eq.~(\ref{bt3}) can be
reexpressed as
\begin{eqnarray}
& &\langle 0|\bar q\sigma^{\mp}_{\hspace{0.2cm}\alpha'}
\gamma_5gG^+_{\hspace{0.2cm}\alpha}h_v|\bar B(v)\rangle
=iF(\mu)\lambda_{\pm}^2(\mu)g_{\alpha'\alpha}^T,\label{sbt3}
\end{eqnarray}
with the normalization factors
$\lambda_+^2\equiv(\lambda_E^2+\lambda_H^2)/2 \approx 0.145$ (0.045)
GeV$^2$ and $\lambda_-^2\equiv (\lambda_E^2-\lambda_H^2)/2\approx
-0.035$ (-0.015) GeV$^2$ from Eq.(\ref{neu}) [Eq.~(\ref{tan})].

The three momenta $P_1-k_1-l_1$, $k_1$ and $l_1$ are assigned to the
initial-state $b$ quark, antiquark, and gluon, respectively.
Equation~(\ref{sbt3}) corresponds to the spin projector
$\gamma^{\pm} \gamma_\alpha^T \gamma_5 \sqrt{2} \lambda_{\pm}^2
/(4y_1)$ from the $B$ meson side,
where the decay constant $f_B$ has been absorbed into the
three-parton $B$ meson distribution amplitude, and the valence gluon
momentum fraction is defined by $y_1=l_1^+/P_1^+$.
For the attachments of the valence gluon in the $B$
meson to the lines in Fig.~\ref{fig1}(a), only the
one to the hard gluon contributes, because of $\gamma^\nu
\gamma^\pm \gamma_\alpha^T \gamma_5\gamma_\nu=0$. All
attachments to the lines in Fig.~\ref{fig1}(b) do not contribute,
since the corresponding Feynman rules have no $m_B$ dependence. As
explained before, the three-parton-to-two-parton amplitudes must be
proportional to $m_B$. Assuming the same three-parton distribution
amplitudes associated with the normalization constants $\lambda_\pm^2$,
we derive
\begin{eqnarray}
H^{3\to 2}&=&-\frac{g^2}{32y_1}\sqrt{2}\lambda_+^2
\frac{tr[(\not k_1 +2 \not l_1)\not
P_2\gamma_\mu \gamma^+]m_b}{[(P_1-k_2)^2-m_b^2](k_1-k_2)^2(k_1+l_1-k_2)^2}
\\ \nonumber &&-\frac{g^2}{32y_1}\sqrt{2}\lambda_-^2
\frac{tr[\not P_2\gamma_\mu \not k_2\gamma^-
]m_b}{[(P_1-k_2)^2-m_b^2](k_1-k_2)^2(k_1+l_1-k_2)^2},\\
&=&\frac{g^2}{4}\frac{1}{y_1}\left(\lambda_-^2
+\frac{x_1+2y_1}{\eta y_2}\lambda_+^2\right)
\frac{P_{2\mu}}{(k_1-k_2)^2(k_1+l_1-k_2)^2}.\label{32}
\end{eqnarray}

It has been known that the form factor $f_1$ is suppressed by
$m_0/m_B$ compared to $f_2$ \cite{TLS}. Therefore, it is natural
that the three-parton contribution corrects only $f_2$ at the
accuracy considered here, which is summarized as
\begin{eqnarray}
f_2^{3p}(q^2)=f_2^{2\to 3}(q^2)+f_2^{3\to 2}(q^2),\label{f3p}
\end{eqnarray}
with the factorization formulas
\begin{eqnarray}
f_2^{2\to 3}&=&\int dx_1dx_2dy_2\int b_1db_1 b_2db_2\phi_B(x_1,b_1)
\Phi_\pi(x_2,y_2)\exp[-s(P_2^-,b_2)]\nonumber\\
& &\times[h_1^{2\to 3}(x_1,x_2,y_2,b_1,b_2)+h_2^{2\to 3}(x_1,x_2,y_2,b_1,b_2)],\\
f_2^{3\to 2}&=&\int dx_1dy_1dx_2\int b_1db_1 b_2db_2\Phi_B(x_1,y_1,b_1,b_2)
\phi_\pi(x_2)\exp[-s(P_2^-,b_1)]h^{3\to 2}(x_1,y_1,x_2,b_1,b_2).
\end{eqnarray}
We have neglected the intrinsic $b$ dependence of the pion
distribution amplitudes, because the suppression of the Sudakov
factor $\exp[-s(P_2^-,b)]$ is strong enough in the large $b$ region
\cite{LS,LY1,KLS,LUY}.
On the contrary, the Sudakov effect associated with the $B$ meson is weak,
since it is dominated by soft dynamics.
For the $B$ meson distribution amplitudes, the intrinsic $b$
dependence is more effective. The hard kernels are written as
\begin{eqnarray}
h_1^{2\to 3}(x_1,x_2,y_2,b_1,b_2)&=&-\frac{\pi}{2}
\frac{2\eta y_2+x_1}{\eta y_2(x_2+y_2)}m_B m_0\alpha_s 
K_{0}\left(\sqrt{x_1(x_2+y_2)\eta}m_Bb_2\right)
\nonumber \\
& &\times \left[\theta(b_1-b_2)K_0\left(\sqrt{x_1x_2\eta}m_B
b_1\right)I_0\left(\sqrt{x_1x_2\eta}m_Bb_2\right)\right.
\nonumber \\
& &\;\;\;\;\left.+\theta(b_2-b_1)K_0\left(\sqrt{x_1x_2\eta}m_Bb_2\right)
I_0\left(\sqrt{x_1x_2\eta}m_Bb_1\right)\right],\nonumber\\
h_2^{2\to 3}(x_1,x_2,y_2,b_1,b_2)&=&-\frac{\pi}{N_c^2}
\frac{m_B m_0}{x_2+y_2}\alpha_s
K_{0}\left(\sqrt{x_1(x_2+y_2)\eta}m_Bb_2\right)
\nonumber \\
& &\times \left[\theta(b_1-b_2)K_0\left(\sqrt{x_1y_2\eta}m_B
b_1\right)I_0\left(\sqrt{x_1y_2\eta}m_Bb_2\right)\right.
\nonumber \\
& &\;\;\;\;\left.+\theta(b_2-b_1)K_0\left(\sqrt{x_1y_2\eta}m_Bb_2\right)
I_0\left(\sqrt{x_1y_2\eta}m_Bb_1\right)\right],\nonumber\\
h^{3\to 2}(x_1,y_1,x_2,b_1,b_2)&=&\frac{\pi}{y_1}\left(\lambda_-^2
+\frac{x_1+2y_1}{\eta y_2}\lambda_+^2\right)\alpha_s
K_{0}\left(\sqrt{(x_1+y_1)x_2\eta}m_Bb_2\right)
\nonumber \\
& &\times \left[\theta(b_1-b_2)K_0\left(\sqrt{x_1x_2\eta}m_B
b_1\right)I_0\left(\sqrt{x_1x_2\eta}m_Bb_2\right)\right.
\nonumber \\
& &\;\;\;\;\left.+\theta(b_2-b_1)K_0\left(\sqrt{x_1x_2\eta}m_Bb_2\right)
I_0\left(\sqrt{x_1x_2\eta}m_Bb_1\right)\right].
\label{dh}
\end{eqnarray}

The functional form of the three-parton $B$ meson distribution amplitude
is still unknown in the literature, though there are already studies of
its relation to the two-parton ones \cite{KKQT,HQW06}.
Below we shall postulate a
simple form for an order-of-magnitude estimate. The involved
two-parton and three-parton meson distribution amplitudes are chosen
as
\begin{eqnarray}
\phi_B(x_1,b_1)& = &N_B f_Bx_1^2(1-x_1)^2
\exp\left[-\frac{1}{2}\left(\frac{x_1m_B}{\omega_B}\right)^2
-\frac{\omega_B^2 b_1^2}{2}\right],\label{os}\\
\Phi_B(x_1,y_1,b_1,b_2)& = &N'_Bf_Bx_1^2(1-x_1-y_1)^2y_1^2
\exp\left[-\frac{\omega_B^2}{2}(b_1^2+b_2^2)\right],\\
\phi_\pi(x_2)&=&6f_\pi x_2(1-x_2)
\left[1+0.44C_2^{3/2}(2x_2-1)+0.25C_4^{3/2}(2x_2-1)\right],
\label{pioa}\\
\Phi_\pi(x_2,y_2) & = & 360\eta_3 f_\pi x_2(1-x_2-y_2)y_2^2
\bigg[1-\frac{3}{2}(7 y_2 -3)\bigg],
\end{eqnarray}
with the parameters $\omega_B=0.4$ GeV \cite{KLS} and $\eta_3=0.015$
\cite{PB2}, and the Gegenbauer polynomials
\begin{eqnarray}
C_2^{3/2}(t)=\frac{3}{2}(5t^2-1)\;,\;\;\;
C_4^{3/2}(t)=\frac{15}{8}(21 t^4 -14 t^2 +1).
\end{eqnarray}
The normalization constants $N_B$ and $N'_B$ are determined through
the relations $\int dx_{1}\phi_B(x_{1},0)=\int
dx_1dy_1\phi_B(x_1,y_1,0,0)=f_B$.
The two-parton $B$ meson and pion distribution amplitudes have been
chosen as in \cite{TLS} in order to have an appropriate
comparison of numerical outcomes.

\begin{figure}[t]
\begin{center}
\includegraphics[width=8.0cm]{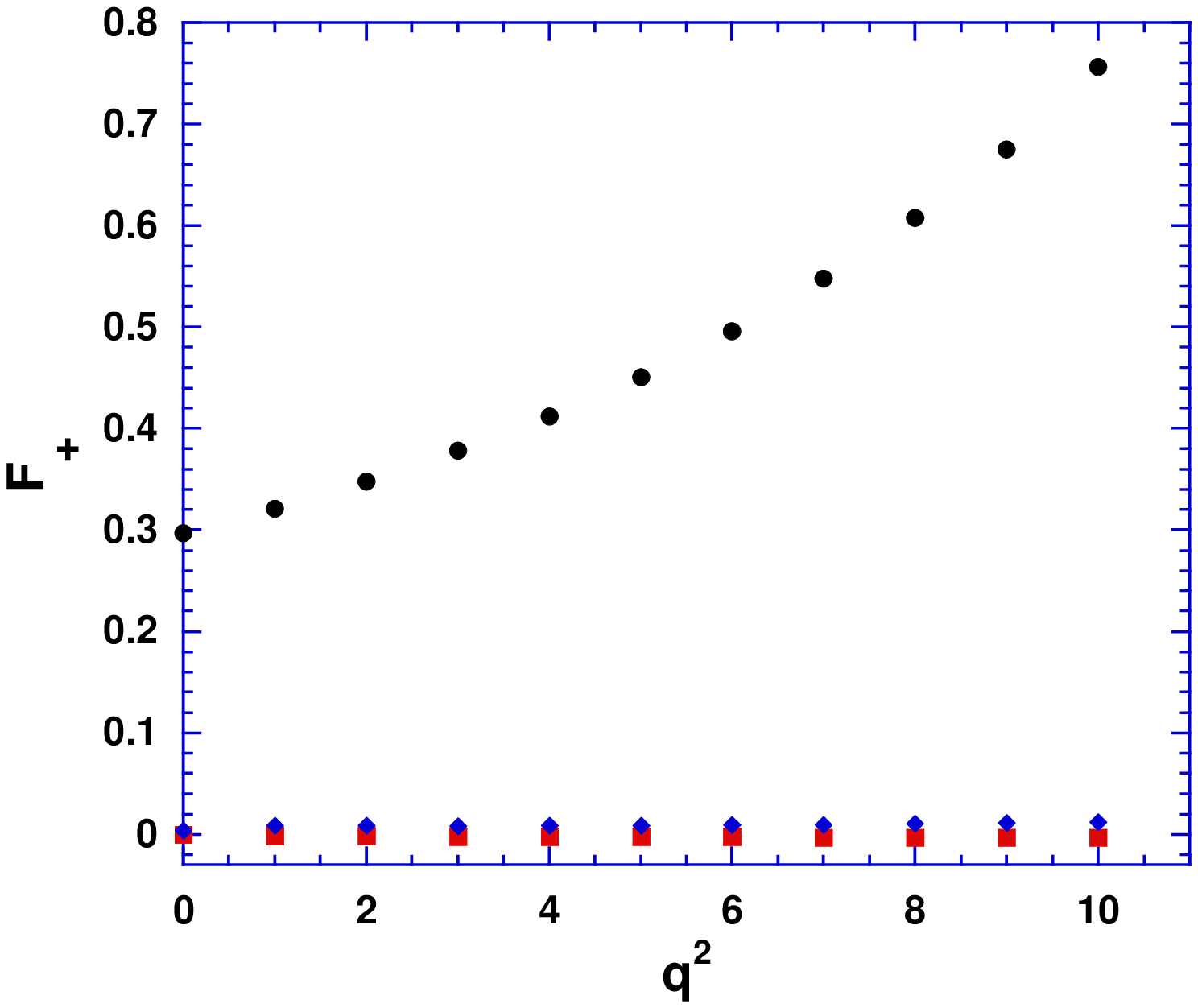}
\includegraphics[width=8.0cm]{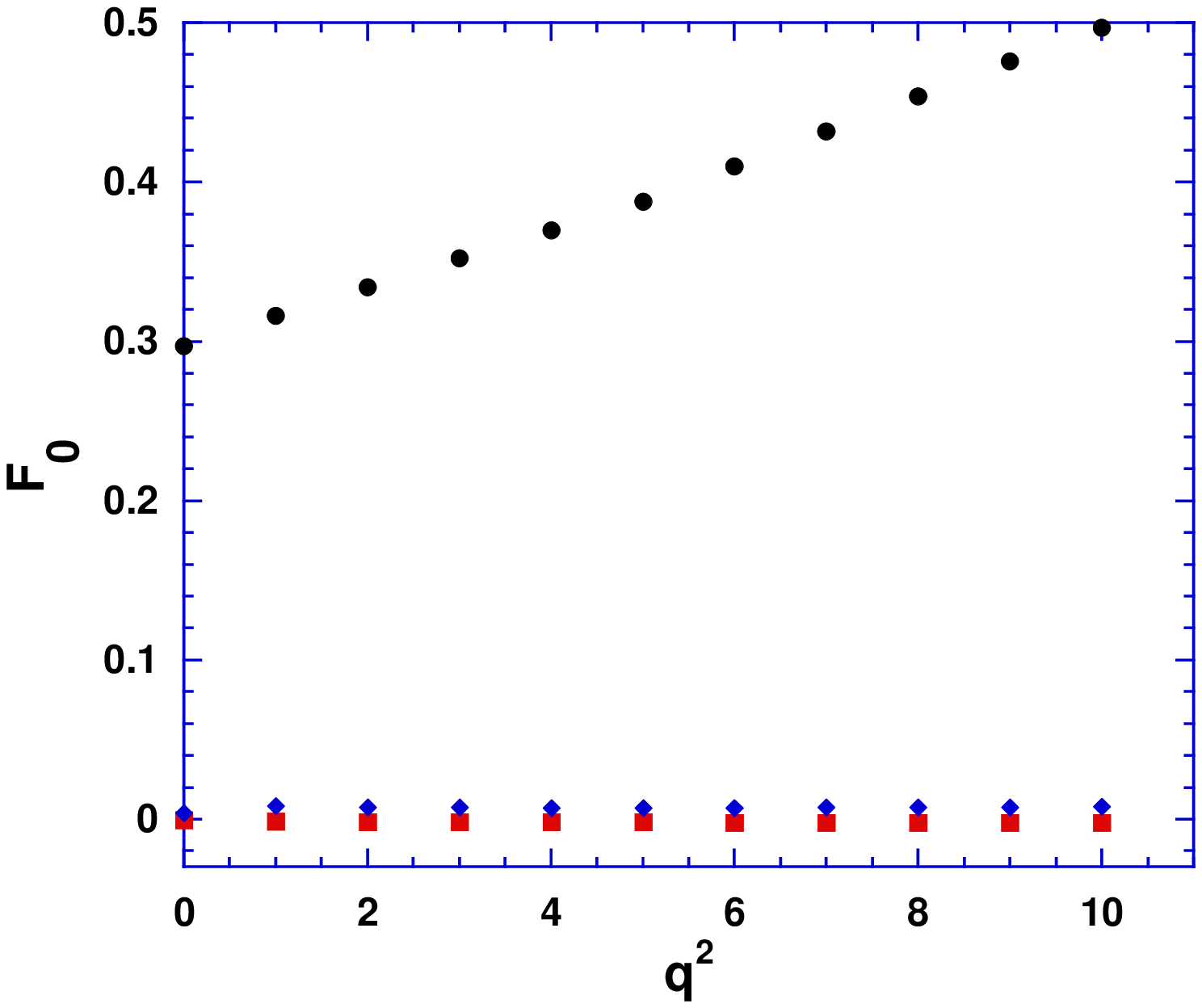}
\caption{Three-parton twist-3 contributions to $F_+(q^2)$ and $F_0(q^2)$
corresponding to Eq.(\ref{neu}) (rhombuses) and to
Eq.~(\ref{tan}) (squares), compared with the leading-power
one (dots) \cite{TLS}.}\label{fig3}
\end{center}
\end{figure}

Equation~(\ref{f3p}) represents the three-parton contribution to the
form factor $f_2$, which then corrects the form factors $F_+$ and
$F_0$ via Eq.~(\ref{f+0}). The numerical results derived from
Eq.~(\ref{f3p}) for $f_B=0.2$ GeV, $f_\pi=0.13$ GeV, $m_B=5.28$ GeV,
$m_0=1.4$ GeV and $\alpha_s=0.5$ are listed in Table~\ref{tbl}, which confirm the
ratio of the three-parton-to-two-parton contribution over the
two-parton-to-three-parton one,
$2\lambda_+^2/(m_Bm_0\eta_3)\approx 2.6$ (0.8) from Eq.(\ref{neu})
[Eq.~(\ref{tan})]. The dominant contribution arises from the
diagrams with the additional valence gluon attaching to the
leading-order hard gluon, i.e., from Eqs~(\ref{c23}) and (\ref{32}).
Figure~\ref{fig3} shows that the three-parton contribution amounts
only up to few percents of the $B\to\pi$ transition form factors
$F_+(0)=F_0(0)\approx 0.3$ at large recoil of the pion. The
relative importance is obvious from the
order-of-magnitude estimate
$\eta_3 m_0/t\sim \lambda_+^2/(m_B t)\sim 1\%$, in which
the scale $\eta_3 m_0$ ($\lambda_+^2/m_B$) is associated with
the spin projector of the three-parton pion ($B$ meson)
distribution amplitude, and $t\sim
1.7$ GeV denotes the characteristic scale involved in $B$ meson
decays at large recoil \cite{TLS}.
Figure~\ref{fig3} also indicates that the three-parton contribution
is of the same order as the third piece in the following
projector associated with the two-parton $B$ meson distribution amplitudes
\cite{BF01,HW05}
\begin{eqnarray}
(\not P_1+m_B) \left[\phi_B(k_1)-\frac{\not n_+-\not
n_-}{\sqrt{2}} {\bar
\phi}_B(k_1)-\Delta(k_1)\gamma^\mu\frac{\partial}{k_{1T}^\mu}\right]\gamma_5,
\label{bwp2}
\end{eqnarray}
with the dimensionless vectors $n_+=(1,0,{\bf 0}_T)$ and
$n_-=(0,1,{\bf 0}_T)$. Collecting the observations obtained in the
literature, we summarize the various contributions to the $B\to\pi$
transition form factors: the first term in the
above projector, which has been considered in \cite{TLS}, gives the
leading contribution. The second term $\bar\phi_B$,
proportional to the difference of the two leading-power
$B$ meson wave functions, contributes 30\% \cite{HW05}. The third term,
proportional to the integration of $\bar\phi_B$ in the momentum fraction,
and the three-parton Fock state contribute only few percents.

\begin{table}
\begin{center}
\begin{tabular}{c||ccccccccccc}\hline\hline
$q^2$
(GeV${}^2$)&0.0&1.0&2.0&3.0&4.0&5.0&6.0&7.0&8.0&9.0&10.0\\\hline
$2\to 3(10^{-2})$ &-0.463&-1.13&-1.237&-1.314&-1.391&-1.472&-1.562&-1.662&-1.774&-1.901&-2.046\\
$3\to 2(10^{-2})$
&1.223&2.885&2.911&2.982&3.091&3.233&3.407&3.614&3.858&4.142&4.476\\\hline
total$(10^{-2})$&0.761&1.754&1.167&1.167&1.700&1.760&1.845&1.952&2.083&2.241&2.429\\\hline\hline
$2\to 3(10^{-2})$ &-0.463&-1.13&-1.237&-1.314&-1.391&-1.472&-1.562&-1.662&-1.774&-1.901&-2.046\\\
$3\to 2(10^{-2})$
&0.376&0.888&0.896&0.918&0.951&0.996&1.049&1.114&1.189&1.277&1.380\\\hline
total$(10^{-2})$&-0.086&-0.243&-0.341&-0.396&-0.439&-0.477&-0.512&-0.549&-0.585&-0.624&-0.667\\\hline\hline
\end{tabular}
\end{center}
\caption{%
Two-parton-to-three-parton and
three-parton-to-two-parton contributions to $f_2(q^2)$ corresponding to
Eq.~(\ref{neu}) (upper half) and to Eq.~(\ref{tan}) (lower half).}
\label{tbl}
\end{table}

\section{CONCLUSION}

In this letter we have extended the investigation of the $B\to\pi$ transition form
factors in the $k_T$ factorization theorem to the three-parton twist-3
level. It was demonstrated that the gauge-dependent pieces cancel
each other in the two(three)-parton-to-three(two)-parton diagrams,
so the gauge invariance of this formalism is verified. The contributions
from the above diagrams were then calculated, and found to be
few percents at most, considering the normalization inputs for the three-parton
$B$ meson distribution amplitudes from QCD sum rules. The
theoretical framework for analyzing three-parton
contributions to $B$ meson decays was established in this work,
which can be compared to other approaches, such as light-cone sum rules
\cite{K01}, the QCD (collinear) factorization \cite{Yeh08}, and
the soft-collinear effective theory \cite{ARS06}.

This work was supported in part by the National Science
Council of R.O.C. under Grant No. NSC-98-2112-M-001-015-MY3, and by
the National Center for Theoretical Sciences.

\end{document}